\documentclass[10pt,twocolumn]{article}

\usepackage{multicol}
\usepackage{graphicx}
\usepackage{amsmath}
\usepackage{lipsum}
\usepackage{color}
\usepackage{siunitx}
\usepackage{authblk}
\usepackage[margin=2cm]{geometry}
\usepackage{mathtools, cuted}

\usepackage{bm}
\usepackage{amsmath}

\title{Coffee-stain growth dynamics on dry and wet surfaces}
\author[1,2,3]{Fran\c{c}ois Boulogne}
\author[1,4]{Fran\c{c}ois Ingremeau}
\author[1]{Howard A. Stone}
\affil[1]{Department of Mechanical and Aerospace Engineering, Princeton University, Princeton, NJ 08544, USA}
\affil[2]{Laboratoire Mati\`ere et Syst\`emes Complexes (MSC), UMR 7057 CNRS, Universit\'e Paris Diderot, B\^atiment Condorcet, 10 rue Alice Domon et Leonie Duquet, Paris, France}
\affil[3]{Laboratoire de Physique des Solides, CNRS, Univ. Paris-Sud, Universit\'e Paris-Saclay, Orsay 91405, France}
\affil[4]{LIPhy, CNRS, and Universit\'e Grenoble Alpes, 140 Rue de la Physique, 38402
Saint-Martin-d'H\`eres, France}

\date{\today}
\begin{document}

\twocolumn[
    \begin{@twocolumnfalse}
        \maketitle
\begin{abstract}
    The drying of a drop containing particles often results in the accumulation of the particles at the contact line.
    In this work, we investigate the drying of an aqueous colloidal drop surrounded by a hydrogel that is also evaporating.
    We combine theoretical and experimental studies to understand how the surrounding vapor concentration affects the particle deposit during the constant radius evaporation mode.
    In addition to the common case of evaporation on an otherwise dry surface, we show that in a configuration where liquid is evaporating from a flat surface around the drop, the singularity of the evaporative flux at the contact line is suppressed and the drop evaporation is homogeneous.
    For both conditions, we derive the velocity field and we establish the temporal evolution of the number of particles accumulated at the contact line.
    We predict the growth dynamics of the stain and the drying timescales.
    Thus, dry and wet conditions are compared with experimental results and we highlight that only the dynamics is modified by the evaporation conditions, not the final accumulation at the contact line.
        \end{abstract}
    \end{@twocolumnfalse}
]

\section{Introduction}

The investigation of the evaporation of a pure liquid drop started in the late 1800s with  papers by Maxwell \cite{Maxwell1877a} and Stefan \cite{Stefan1882}, and continued in 1918 with a further contribution by Langmuir \cite{Langmuir1918}, for the specific case of a spherical drop suspended in air.
Also, the evaporation of disk-shaped liquid drops has been motivated by the understanding of plant transpiration \cite{Thomas1917,Thomas1917a,Jeffreys1918} and to predict the lifetime of droplets \cite{Houghton1933,Stauber2014,Stauber2015a}.
The addition of small particles to the liquid drop leaves, upon evaporation, a ring stain along the contact line, as commonly observed for drops of tea or coffee.
This so-called coffee-stain effect was first investigated about 20 years ago by Deegan \textit{et al.} \cite{Deegan1997,Deegan2000a}.
It is now well-established that the accumulation of the particles at the contact line is the result of an outward radial flow owing to evaporation \cite{Deegan1997,Deegan2000,Berteloot2008,Pham2010,Berteloot2012a}.
This accumulation pins the contact line and the loss of the volatile solvent decreases the drop height \cite{Hu2002a,Hu2005}.
Consequently, to replenish the loss of solvent at the pinned contact line, the liquid flows radially.
The characteristics of this flow transporting the particles in the drop depends on the flow field in the drop and on the details of the evaporation profile.

The laminar radial flow field commonly observed in simple volatile liquids can be modified by changing the composition of the liquid.
Thermal or solutal Marangoni effects have been extensively studied in the literature, as they can contribute to a chaotic flow in a binary solvent \cite{Kim2016} or a reverse flow from surfactant concentration gradients \cite{Kim2016}.
Also, instead of evaporating the liquid from the liquid-vapor interface, the absorption by the substrate is known to modify the particle deposition to achieve a spatially more uniform coating \cite{Boulogne2015b}.

For simple liquids characterized by a laminar radial flow, the transport of particles suspended in a evaporating drop depends solely on the evaporative flux.
To describe the particle accumulation at the contact line, it is necessary to establish first the evaporative flux, which is function of the solvent diffusion in the gas phase, and the shape of the liquid-vapor interface \cite{Deegan1997,Deegan2000,Eggers2010,Marin2011,Gelderblom2012}.
Mathematically, it is established that for a sessile drop on a non-evaporating surface, the evaporative flux at the liquid-vapor interface diverges at the contact line \cite{Deegan1997,Deegan2000,Eggers2010}.
By using the analogy between diffusive concentration fields and electrostatic potential fields, the divergence of the evaporative flux can be understood as a tip-shape effect \cite{Picknett1977,Deegan2000a}.

Therefore, the evaporation profile is well-established for a single drop on a dry surface.
However, in many common situations and industrial settings, drops in the presence of other surrounding droplets, a liquid film or a porous substrate imbibed with volatile liquid are also subject to evaporation.
For instance, the mutual influence of drops \cite{Sokuler2010,Laghezza2016,Carrier2016} can be crucial for sprayed surface coatings where deposited droplets are closely spaced.
As a result,  the vapor concentration field around a drop can vary substantially in the presence of a drying environment.
In particular, we can expect the evaporative flux at the surface of such drops to be more homogeneous.
Thus, as the liquid flow is a function of the evaporation and as it transports suspended particles \cite{Popov2005,Zheng2009,Monteux2011,Kaplan2015}, the resulting number of particles accumulated at the contact line may be expected to depend on the environmental conditions.
To the best of our knowledge, a detailed comparison between a spatially homogeneous flux of solvent and a diverging flux has not been reported to date.

In previous studies \cite{Boulogne2015b,Boulogne2016b}, we focused our attention on the absorption of a drop containing micrometer size particles by a swelling hydrogel.
Instead of absorption, our present study is motivated by the drying of a drop containing microparticles on a wet hydrogel.
In such a configuration, the drop and the gel evaporate.
Therefore, we expect that the evaporation of the drop is modified by these new environmental conditions and that the transport of colloidal particles is affected.

In this paper, we investigate theoretically  the coffee stain effect of a drop  placed in the center of an evaporating surface that is much larger than the drop diameter and we illustrate the concepts with experimental observations.
First, we argue theoretically that the divergence of the flux at the contact line is strongly reduced in this geometry compared to the classic sessile drop on a dry surface.
Our study focuses on dilute suspensions, such that particles are expected to act as tracers and  do not disturb the velocity field.
However, their presence at the contact line helps to maintain a constant radius of the drop and we ignore the second regime where the contact line eventually recedes.
The main result of this paper concerns the time evolution of the particle density close to the contact line, which is different in the two cases of a single drop on a non-evaporating surface and a drop on a drying surface.
This observation supports the suppression of the diverging flux at the contact line.
Nevertheless, we show that for both conditions, the particles are transported toward the contact line such that the final patterns are similar.
Therefore, we conclude that the divergence of the evaporating flux at the contact line for a drop on a dry surface is not crucial to obtain coffee stain effect.
Based on the model and the experimental results, the coffee stain mechanism is discussed in section \ref{sec:observation}.

The outline of the paper is organized as follow.
In Section \ref{sec:model}, we first recall the evaporation dynamics of a single drop on a non-evaporating surface.
Then, we derive the evaporation dynamics of a large circular hydrogel to show that  a small drop placed in the center of this gel evaporates with a nearly uniform flux.
Furthermore, we calculate the velocity fields in both cases as well as the temporal evolution of the number of particles accumulated at the contact line.
In Section \ref{sec:experiments}, we present our experimental setup for the preparation of the dry and wet configurations.
Then, we compare our experimental results to the theoretical predictions regarding the drop lifetime and the time evolution of the number of particles accumulated at the contact line.

%%%%%%%%%%%%%%%%%%%%%%%%%%%%%%
%
% MODEL
%
%%%%%%%%%%%%%%%%%%%%%%%%%%%%%%

\section{Model}\label{sec:model}

\begin{figure*}
    \centering
    \includegraphics[height=3.4cm]{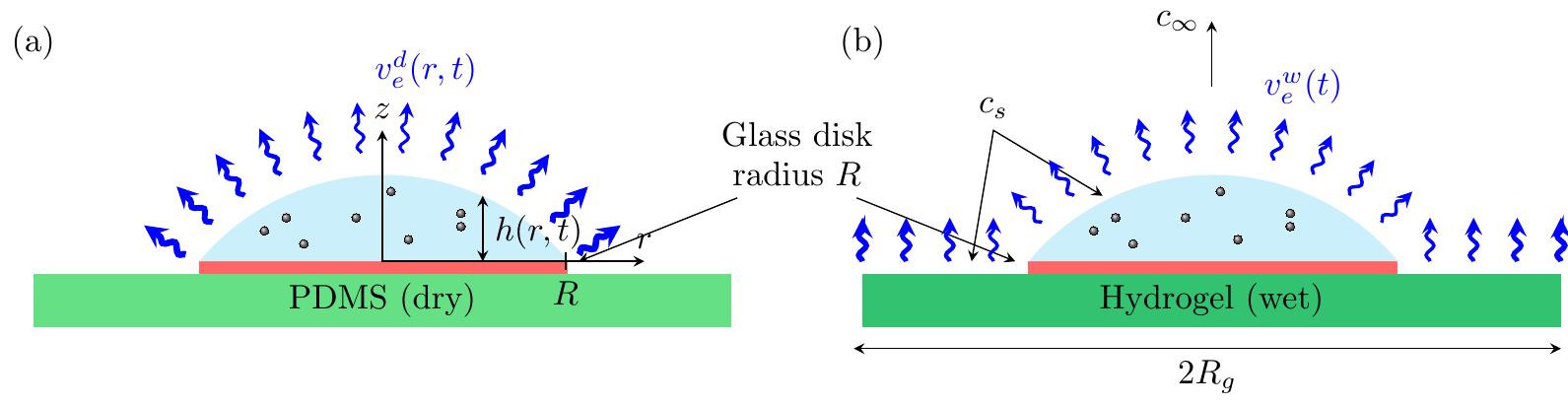}
    \caption{
        Sketches of the (a) dry and (b) wet configurations that also illustrate the materials used experimentally to achieve both cases.
        Each drop is sitting on a thin glass disk of radius $R$ and the profile of the drop interface is denoted $h(r,t)$.
        The thin glass disk prevents a solvent flux between the drop and the substrate and ensures that the surface conditions are the same in both cases for the particles in the drop.
        The vapor concentrations at saturation and at infinity are denoted $c_s$ and $c_\infty$, respectively.
        The thickness of arrows indicates qualitatively the strength of the evaporative flux.
    }\label{fig:situation}
\end{figure*}

\subsection{Drop shape}
We consider two distinct situations depicted in Figure \ref{fig:situation} where a liquid drop evaporates on a substrate.
We assume that the geometry is axisymmetric and we use cylindrical $(r,z)$ coordinates.
The capillary length is $\ell_c = \sqrt{\gamma/(\rho g)}$, where $\gamma$ is the surface tension, $g$ is the gravitational acceleration and $\rho$ the liquid density.
We assume that the contact line at $r=R$ is pinned during the evaporation, which is consistent with the experiments where we observe that the drop has a constant radius.
This defines the constant radius evaporation mode and the drop height decreases in time.
Nevertheless, the contact line eventually recedes at the very end of the drying process, when the liquid film height is typically of the order of a few times the particle size.

For a drop radius $R < \ell_c$, to a good approximation the drop shape is a spherical cap.
Thus, at each time, the drop profile $h(r,t)$ (Figure \ref{fig:situation}) is described by a portion of a sphere,

\begin{equation}
    h(r,t) = \sqrt{ \left(  \frac{R^2 + h_0(t)^2}{2h_0(t)} \right)^2 - r^2 }  - \frac{(R^2 - h_0(t)^2)}{2h_0(t)},\label{eq:HOMO_drop_shape_general}
\end{equation}
with $h_0(t) = h(0,t)$.
This solution ensures that the curvature is only a function of time, such that the capillary pressure is uniform in the drop and capillary effects alone do not generate any flow.

We assume that the drop height is much smaller than the drop radius, $h_0\ll R$, which corresponds to small contact angles.
This assumption is necessary in the later sections to apply the lubrication approximation to develop analytical results.
Therefore, the drop shape is approximated by
\begin{equation}
    h(r,t) \simeq h_0(t)\, \left( 1 - \frac{r^2}{R^2} \right) .\label{eq:HOMO_drop_shape}
\end{equation}

With the assumption of a pinned contact line ($\textrm{d}R/\textrm{d}t = 0$), the time derivative of the drop height $h(r,t)$ defined by equation (\ref{eq:HOMO_drop_shape}) is given by

\begin{equation}
    \frac{\partial h}{\partial t} =  \left( 1 - \frac{r^2}{R^2} \right) \,\frac{{\rm d}h_0}{{\rm d} t}. \label{eq:HOMO_dhdt_1}
\end{equation}
The drop has the volume of a spherical cap $\Omega(t) =\frac{\pi}{2} h_0(t) \left(\frac{h_0(t)^2}{3}+R^2\right)\simeq \pi R^2 h_0(t) / 2$.
We define $Q_e(t)$ as the total evaporative flux given by $Q_e(t) = \int_S v_e \mbox{d}S$, where $S$ designates the liquid-vapor interface of the drop and $v_e$ denotes the local evaporation speed, \textit{i.e.} the volume of liquid that evaporates per unit area per unit time.
The time variation of this volume corresponds to the loss of liquid by evaporation, \textit{i.e.} ${\rm d} \Omega / {\rm d} t = - Q_e(t)  $.
Therefore, equation (\ref{eq:HOMO_dhdt_1}) simplifies to
\begin{equation}
    \frac{\partial h}{\partial t} = -\left( 1 - \frac{r^2}{R^2} \right) \frac{2 Q_e(t)}{\pi R^2}.\label{eq:HOMO_temporal_derivative}
\end{equation}
In the next section, we present the evaporation dynamics in two different environmental situations.

%%%%%%%%%%%%%%%%%%%%%%%%%%%%%%
% evaporation
%%%%%%%%%%%%%%%%%%%%%%%%%%%%%%

\subsection{Diffusion-limited evaporation}

We aim to analyze the difference between drops evaporating on a dry surface and on a surface that presents an evaporative flux of the same liquid as the drop.
To do so, we need to calculate the evaporation speed in these distinct cases.
First, we present the well-known equations for diffusion-limited evaporation.
Then, we establish the vapor concentration surrounding a disk of gel to define the conditions for which the evaporative flux can be considered to a good approximation as homogeneous close to its center.
Finally, we give the expressions of the evaporative flux for a single drop on a dry surface and for a drop surrounded by an evaporating surface.

\subsubsection{Theoretical background}

We denote by $c_v$ the vapor mass concentration in the gas surrounding the evaporating liquid.
For diffusion-limited evaporation \cite{Picknett1977,Poulard2005a}, the temporal and spatial evolution of the vapor concentration follows
\begin{equation}
    \frac{\partial c_v}{\partial t} = {\cal D} \nabla^2 c_v,\label{eq:fick}
\end{equation}
where ${\cal D}$ is the diffusion coefficient of the vapor in the gas phase.

We consider the characteristic length scale ${\cal L}$ of the vapor concentration gradient, which is related to the geometry of the system at long timescales \cite{Sultan2005,Boulogne2015a}.
We also  denote ${\cal V}_e$ as the typical value of the evaporation speed $v_e$.
In particle-tracking experiments reported below, the relative effect of the diffusive time of the vapor in the gas phase $\tau_D = {\cal L}^2/{\cal D}$ and the evaporation time $\tau_{e}={\cal L}/ {\cal V}_e$ defines the P\'eclet number:
\begin{equation}
    \textrm{Pe} = \frac{{\cal V}_e \, {\cal L}}{{\cal D}}.\label{eq:Peclet_number}
\end{equation}
The appropriate choice of the length scale ${\cal L}$ is made in the next sections for each geometry.
The diffusion coefficient ${\cal D}$ for water vapor in air at room temperature is ${\cal D} = 2 \times 10^{-5}$ m$^2$/s.
In our experimental conditions, the typical evaporation speed for water is ${\cal V}_e\approx 10^{-8}$ m/s (see Appendix).
If the P\'eclet number is small, then equation (\ref{eq:fick}) can be simplified to the Laplace equation

\begin{equation}
    \nabla^2 c_v = 0.\label{eq:laplace_equation}
\end{equation}
The evaporation velocity at the liquid-vapor interface is \cite{Cazabat2010}
\begin{equation}
    v_e(r,t) =   - \frac{{\cal D}}{\rho} {\bf n} \cdot {\bm\nabla} c_v \qquad \mbox{at} \quad z=h(r,t) \label{eq:gradvap}
\end{equation}
where $\rho$ is the liquid density, $h(r,t)$ is the vertical position of the liquid-vapor interface and ${\bf n}$ the unit normal vector directed into the gas phase.

    \subsubsection{Dry configuration}
    
    Under the assumption of diffusion-limited evaporation, we recall the evaporation dynamics of a sessile drop of radius $R_d$ (Figure \ref{fig:situation}a).
    The characteristic length scale ${\cal L}$ is the glass disk radius, thus $\textrm{Pe} \ll 1$.
    For a contact angle $\theta_c$, the evaporation speed is given by \cite{Deegan1997,Deegan2000a}

    \begin{equation}
        v_e^d(r) = \frac{2{\cal D} (c_s - c_\infty)}{\pi \rho}   \left\{
            \begin{array}{ll}
                \frac{f(\lambda)}{(R^2 - r^2)^\lambda} & \mbox{for } r < R, \\
                0 & \mbox{for } r > R,
            \end{array}
            \right.\label{eq:DEE_current}
        \end{equation}
        where $\lambda = \frac{\pi - 2\theta_c}{2\pi - 2\theta_c}$ and $f(\lambda)$ is a known function.
        For $\theta_c \ll 1$, we have $\lambda= 1/2$ and $f(\lambda)=1$ \cite{Deegan2000a,Eggers2010}, where we recover the particular solution of a flat disk of radius $R$ given by equation (\ref{eq:HOMO_current}).
        Consequently, the total evaporative flux of the drop in this dry configuration is
        \begin{equation}
            Q_e^d = \frac{ 4 {\cal D} (c_s - c_\infty) R}{\rho} .\label{eq:DEE_Q}
        \end{equation}
        From equation (\ref{eq:HOMO_temporal_derivative}) and $Q_e^w$ from equation (\ref{eq:DEE_Q}), which is time independent, the height at the center of the drop varies according to
               \begin{equation}\label{eq:DEE_height_time}
            h_0(t) = h_i \left(1-\frac{t}{\tau_e^d}\right),
        \end{equation}
where the initial drop height is $h_i = h(r=0, t=0)$
      and $\tau_e^d = \frac{\rho h_i R \pi}{8{\cal D} (c_s - c_\infty)}$ is the evaporation time.

% WET
\subsubsection{Wet configuration}\label{subsec:wet}

Next, the evaporation dynamics of the wet configuration represented in Figure \ref{fig:situation}(b) is investigated.
We consider a circular slab of hydrogel of radius $R_g$, without the glass disk and the drop.
The typical length scale for the vapor concentration gradient is the gel radius $R_g$ and, for example, for ${\cal L}=R_g=1.75$ cm, we have $\textrm{Pe}\ll 1$.
This geometry corresponds to the particular case $\theta_c=0$ in equation (\ref{eq:DEE_current}).
The vapor concentration field can be simply written as \cite{Lebedev1965,Marder1981}
\begin{equation}\label{eq:vapor_disk}
    {\cal C}_v^w(\kappa, \sigma) =  \frac{c_v^w(\kappa, \sigma) - c_\infty }{(c_s - c_\infty)} =  1- \frac{2}{\pi} \arctan(\sigma),
\end{equation}
where we use the oblate spheroidal coordinates ($\kappa, \sigma$) defined as $r^2 = R_g^2(1-\kappa^2)(1+\sigma^2)$ and $z= R_g\kappa\sigma$.
The evaporation velocity at $z=0$ is given by
\begin{equation}
    v_e^w(r) = \frac{2{\cal D} (c_s - c_\infty)}{\pi \rho}
    \left\{
        \begin{array}{ll}
            \frac{1}{\sqrt{R_g^2 - r^2}} & \mbox{for } r < R_g, \\
            0 & \mbox{for } r> R_g,
        \end{array}
        \right.\label{eq:HOMO_current}
    \end{equation}
which is equivalent to equation (\ref{eq:DEE_current}) with $R$ replaced by $R_g$ and $\theta_c=0$.

\begin{figure*}
        \centering
        \includegraphics[height=4.5cm]{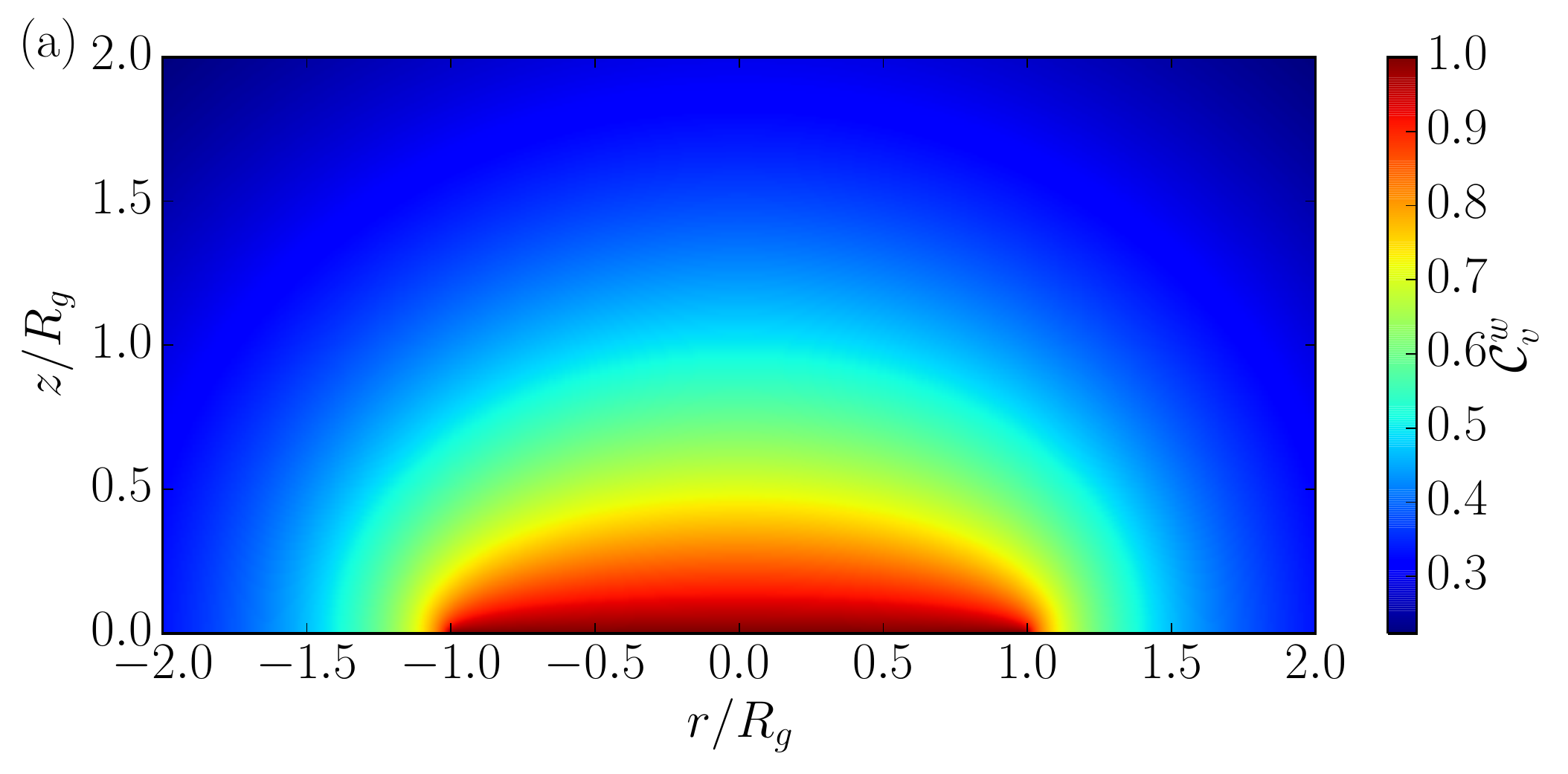}\\
        \includegraphics[width=.8\linewidth]{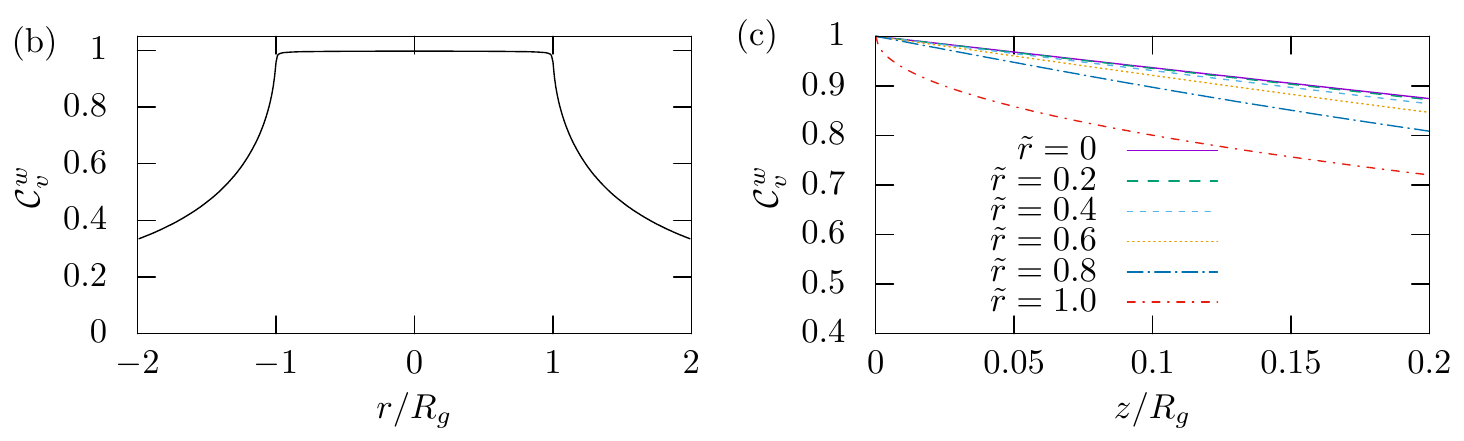}\\
        \caption{Vapor concentration obtained from equation (\ref{eq:vapor_disk}).
            (a) Vapor concentration map surrounding a drying disk of radius $R_g$ obtained from equation (\ref{eq:vapor_disk}).
            (b) Radial vapor concentration profile at $z/R_g=0.1$.
            (c) Vertical vapor concentration profiles at different radial positions $\tilde r = r/R_g$.
        }\label{fig:drying_disk}
    \end{figure*}

    The representation of the vapor concentration surrounding the hydrogel disk is shown in Figure \ref{fig:drying_disk}(a) and we focus our interest in the center of the circular disk of radius $R_g$ where the drop will be placed, \textit{i.e.} $r/R_g < 0.1$.
        Near the edge of the slab, we expect a divergence of the evaporative flux as observed for a sessile drop.
    As is clear from Figures \ref{fig:drying_disk}(b-c), the vapor concentration and its vertical gradient
    above the center of a flat circular slab only weakly depends on the radial and vertical coordinates.
This makes the hydrogel surface a good candidate to study the evaporation of a small droplet exposed to a homogeneous evaporative flux near the center of the gel disk.

    As a consequence of the results shown in Figure \ref{fig:drying_disk}, the drying velocity $v_e^w$ has a weak variation for $r\in [0,R]$ if $R\ll R_g$.
    Therefore, we will consider that in the configuration similar to the one represented in Figure \ref{fig:situation}(b), but with $R < 0.1 R_g$, the drying velocity of the drop is uniform and is equal to
    \begin{equation}
        v_e^w = \frac{2{\cal D} (c_s - c_\infty)}{\pi \rho  R_g},
        \label{eq:HOMO_avvelocity}
    \end{equation}
    where $\frac{{\cal D} (c_s - c_\infty)}{\rho } = 2\times 10^{-10}$ m$^2$/s in our experimental conditions.
    Thus, for an air-vapor interfacial area approximated as $\pi R^2$, the evaporative flux from the drop is
    \begin{equation}
        Q_e^w = v_e^w \,\pi R^2 = \frac{2{\cal D} (c_s - c_\infty) R^2}{\rho R_g}, \label{eq:HOMO_Q}
    \end{equation}
    considering that in the wet configuration, the local evaporative flux is uniform.
            As expected on physical grounds, in the present experimental conditions, the average evaporation velocity for the dry configuration $\bar{v}_e^d = Q_e^d / ( \pi R^2) > v_e^w$ since $R<R_g$.

		Similarly to equation (\ref{eq:DEE_height_time}), the height at the center of the drop follows
       \begin{equation}\label{eq:HOMO_height_time}
            h_0(t) = h_i \left(1-\frac{t}{\tau_e^w}\right),
        \end{equation}
         where 
        $\tau_e^w = \frac{h_i }{2 v_e^w}$ is the evaporation time.

        \subsection{Convective evaporation}\label{subsec:convection}

        Our previous derivations assume that the evaporation is limited by diffusion.
        In this section, we briefly analyze the effect of convection on evaporation.
        Recent studies have addressed the importance of convection on the evaporation of drops \cite{Shadizadeh-Bonn2006,Weon2011,Kelly-Zion2011,Kelly-Zion2013,Carle2013a,Somasundaram2015}.
        In the case of water, humid air is less dense than dry air, which leads to natural convection in the surrounding air.

         In order to close the problem, the density of air as to be expressed as a function of the water concentration, which is related to the partial pressure of vapor. We denote $p_d$ and $p_v$ the partial pressures of dry and humid air, respectively.
         The density of air is then
        \begin{equation}
            \rho_m =  \frac{p_{d}M_{d}+p_{v}M_{v}}{{\cal R} T},
        \end{equation}
        where $M_{d}$ and $M_{v}$ are the molar masses of dry and wet air, respectively, ${\cal R}$ is the ideal gas constant and $T$ is the absolute temperature.
        % ( 0.028964 kg/mol and 0.018016 kg/mol).

\begin{figure}
            \centering
            \includegraphics[width=.99\linewidth]{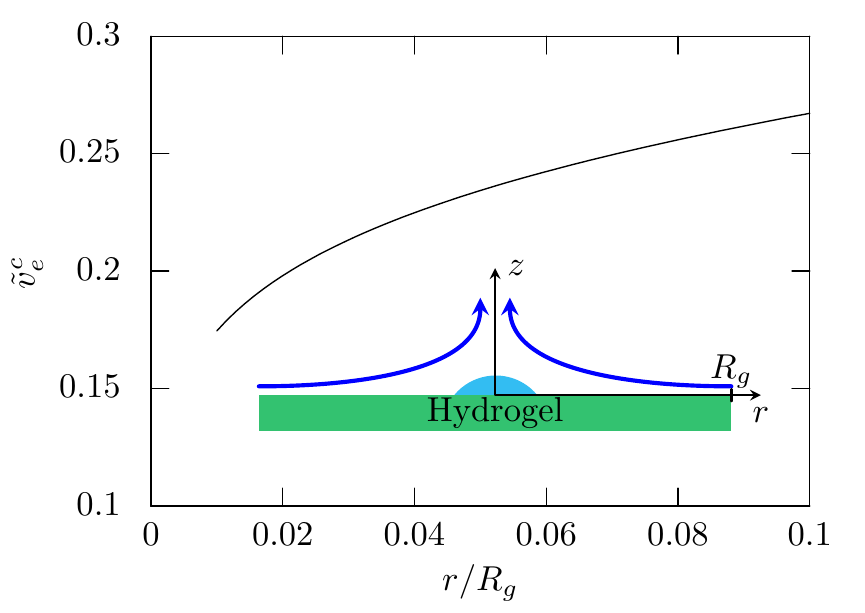}
            \caption{
                Free-convection driven evaporative flux $\tilde v_e^c$ near the center of the gel disk along the radius for ${\rm Sc}=0.8$ and ${\rm Gc} = 6 \times 10^4$.
                The radial range corresponds to the droplet position ($R/R_g \approx 0.1$).
                The sketch introduces the notations with the approximate flow field indicated.
            }\label{fig:merkin}
        \end{figure}

        Therefore, we now consider the steady state convectively driven evaporation above a circular disk of radius $R_g$ with a boundary-layer flow and use the Boussinesq approximation (Figure \ref{fig:merkin}) \cite{Rotem1969,Traugott1975}.
        The convective flow is driven by the density difference between humid air located near the liquid-vapor interface and the ambient air.
        We denotes $(u_r, u_z)$ as the velocity field in the vapor phase in cylindrical coordinates.
        The continuity equation is
        \begin{equation}\label{eq:conv_continuity}
            \frac{1}{r}  \frac{\partial (r u_r)}{\partial r} + \frac{\partial u_z}{\partial z} =0
        \end{equation}
        and in the steady state, the radial momentum equation is
        \begin{equation}\label{eq:conv_momentum}
            u_r \frac{\partial u_r}{\partial r} + u_z \frac{\partial u_r}{\partial z} = - \frac{1}{\rho_m} \frac{\partial p}{\partial r} + + \nu \nabla^2 u_r,
        \end{equation}
        where $\nu$ is the kinematic viscosity of vapor and $\rho_m$, the vapor phase density, considered as a constant in the framework of the Boussinesq approximation.
        The vertical momentum equation is written,
        \begin{equation}\label{eq:conv_momentumz}
            u_r \frac{\partial u_z}{\partial r} + u_z \frac{\partial u_z}{\partial z} = - \frac{1}{\rho_m} \frac{\partial p}{\partial z} + \nu \nabla^2 u_z.
        \end{equation}
        The vertical pressure gradient is given by
        \begin{equation}\label{eq:conv_pressure}
            \frac{\partial p}{\partial z} = (\rho_m - \rho_\infty) g
        \end{equation}
        where $\rho_\infty$ is the air density far from the disk.
        This gradient is the driving force for the convection, which induces a vertical motion described by equation (\ref{eq:conv_momentumz}).
        The steady state vapor mass conservation satisfies
        \begin{equation}\label{eq:conv_diffusion}
            u_r \frac{\partial c_v}{\partial r} +  u_z \frac{\partial c_v}{\partial z} = {\cal D} \frac{\partial^2 c_v}{\partial z^2}.
        \end{equation}
        On the surface of the disk, the boundary conditions are $u_r(r,z=0) = u_z(r,z=0)=0$ and $c_v(r,z=0)=c_s$.
        Far from the disk, $u_r(r,z\rightarrow\infty) = 0$  and $c_v(r,z\rightarrow\infty) = c_\infty$.

        The mass Grashof number compares the buoyancy forces to the viscous forces represented, respectively, by the first and the second terms on the right-hand side of both equations (\ref{eq:conv_momentum}) and (\ref{eq:conv_momentumz}),
        \begin{equation}\label{eq:grashof_number}
            {\rm Gc} = \left| \frac{\rho_s - \rho_\infty}{\rho_\infty} \right| \left( \frac{g R_g^3}{\nu^2} \right),
        \end{equation}
        where $\rho_s$ and $\rho_\infty$ are, respectively, the vapor density at saturation and at infinity.
        For ${\rm Gc} \gg 1$, we expect that convection takes place above the evaporating surface and that the diffusion-limited solution given by equation (\ref{eq:vapor_disk}) is not valid.
        As the mass Grashof number is proportional to $R_g^3$, convection must be important for large evaporating surfaces or very volatile liquids \cite{Dehaeck2014}.
        For instance, for conditions typical of our experiments with a circular gel of radius $R_g = 1.75$ cm, setting the humidity of the glove box $c_\infty = 0.5 c_{s}$ leads to ${\rm Gc} =  10^3$.
        The Schmidt number $Sc=\nu / {\cal D}$ describes the relative thickness of the velocity and the vapor concentration boundary layers and appears in the set of equations (\ref{eq:conv_continuity}-\ref{eq:conv_diffusion}) to be solved.
        For air at room temperature, the kinematic viscosity is $\nu = 1.6\times 10^{-5}$ m$^2$/s and ${\rm Sc} = 0.8$.

        Since we are interested in the evaporative flux of a drop placed  at the center of the gel, we focus on the evaporation rate of the gel at this location.
        To solve equations (\ref{eq:conv_continuity}-\ref{eq:conv_diffusion}), we follow the numerical scheme presented by \cite{Merkin1983} for the free convection above the center of a heated horizontal circular disk.
        In this numerical scheme, the absolute value of $\tilde v_e^c(\tilde r)$ cannot be predicted by this numerical scheme.
        Nevertheless, we obtain the typical variation of the evaporative flux for small radii where we place a drop in our experiments.
        In Figure \ref{fig:merkin}, we show that the evaporative flux increases along the radius more significantly than for the diffusion-limited situation.

        From this analysis, we expect that the evaporative flux of a drop placed in the center of the gel does not diverge at the contact line as a drop in a dry configuration.
        In the following, as a first approximation, we will assume that the evaporative flux is uniform in spite of the presence of convection above the hydrogel disk.
        Thus, in this wet configuration, we will consider that convection only increases the evaporation rate compared to the diffusion-limited situation.
        So, the characteristics of the flow in the drop resulting from the evaporation is independent of the Grashof number, which affects only the evaporation rate and the flow speed.

        %%%%%%%%%%%%%%%%%%%%%%%%%%%%%%
        % Hydro
        %%%%%%%%%%%%%%%%%%%%%%%%%%%%%%

        \subsection{
            Evaporation of dilute colloidal drops
        }
        In the previous section we characterized the velocity of the vapor at the liquid-gas interface for both dry and wet configurations.
        Since we are interested in the motion of the colloidal particles in the drop and more specifically in the formation of particle deposits, we need to describe the flow field in the drop during the constant radius evaporation mode.

        In the following, we assume that the colloidal suspension is dilute so that any accumulation does not affect the  drop shape and the flow field; also, hydrodynamic interactions are neglected and particles are assumed to follow the streamlines.
        \subsubsection{Hydrodynamics}

        Let $(v_r, v_\theta, v_z)$ denote the velocity field in the liquid in cylindrical coordinates $(r,\theta,z)$.
        The problem is invariant by rotation around the $z$ axis.
        Thus, $v_\theta = 0$, $\partial /\partial \theta = 0 $, and we only consider $v_r(r, z, t)$ and $v_z(r, z, t)$.
        Under this assumption, the continuity equation is
        \begin{equation}
            \frac{1}{r} \frac{\partial (r v_r)}{\partial r} + \frac{\partial v_z}{\partial z} = 0.\label{eq:HOMO_continuity}
        \end{equation}

        Also, the Navier-Stokes equation, analyzed within the lubrication approximation, $h_0\ll R$ \cite{Batchelor2000}, provides a linear stress balance along the $r$-direction
        \begin{equation}
            \eta \frac{\partial^2 v_r}{\partial z^2} =  \frac{\partial p}{\partial r},
        \end{equation}
        where $\eta$ is the viscosity and $p$ is the pressure.
        A first integration of this equation with respect to the variable $z$ gives
        \begin{equation}
            \frac{\partial  v_r}{\partial z } =  \frac{1}{\eta}  \frac{\partial p}{\partial r}  \left(z-h(r,t)\right)
        \end{equation}
        where we used the boundary condition of a zero stress at the liquid-vapor interface $\frac{\partial  v_r}{\partial z } (z=h, t) = 0$.
        With the second boundary condition of a zero tangential velocity on the substrate $v_r(z=0, t) = 0$, a second integration leads to a Poiseuille flow
        \begin{equation}
            v_r(r,z,t) =  \frac{1}{\eta}  \frac{\partial p}{\partial r}  \left(\frac{z^2}{2}-h(r,t)z\right).\label{eq:HOMO_radial_velocity_gradp}
        \end{equation}
        We define the average velocity $\bar{v}_r(r, t)$ over the drop height $h(r, t)$
        \begin{eqnarray}
            \bar{v}_r(r,t) =  \frac{1}{h(r,t)} \int_0^{h(r,t)} v_r(r,z,t) \, {\rm d}z = - \frac{h(r,t)^2}{3\eta}  \frac{\partial p}{\partial r}.\label{eq:A}
        \end{eqnarray}
        Substituting equation (\ref{eq:A}) in (\ref{eq:HOMO_radial_velocity_gradp}), we have
        \begin{equation}
            v_r(r, z, t) =  \frac{3 \bar{v}_r(r,t) }{2 h(r,t)^2} (2h(r,t) z - z^2).\label{eq:HOMO_radius_velocity_step1}
        \end{equation}

        To obtain the average velocity $\bar{v}_r(r, t)$ as a function of the evaporative flux of solvent, we use the local continuity equation obtained by a mass conservation in a slice between $r$ and $r+{\rm d}r$, which yields
                    \begin{equation}\label{eq:bilan_local_end}
                        \frac{\partial h}{\partial t} + \frac{1}{r}  \frac{\partial (h r \bar{v}_r )}{\partial r} + v_e(r) =0,
            \end{equation}
        where the time derivative of the liquid height is given by equation (\ref{eq:HOMO_temporal_derivative}).
        In the next sections, we solve this equation for two distinct evaporating conditions.

        \subsubsection{Uniform evaporation for a wet surface}

        Let us consider first the wet configuration.
        We assume that the drop dries with a velocity $v_e^w$ independent of the radial coordinate $r$ and constant in time.
        This assumption is supported by the analysis made in section \ref{subsec:wet} and \ref{subsec:convection} for the diffusive and the convective regimes respectively.
        By integrating equation (\ref{eq:bilan_local_end}) with respect to $r$, we have
        \begin{equation}\label{eq:HOMO_vr_bar_final}
            \bar{v}_r^w(r,t) = \frac{r\,v_e^w}{2h(r,t)}   \left( 1-\frac{r^2}{R^2} \right) = \frac{r v_e^w}{2h_0(t)},
        \end{equation}
        where we used the condition $\bar{v}_r^w(r=0,t)=0$.
        Hence, $\bar{v}_r^w$ is a linear function of $r$.
        Consequently, the radial velocity $v_r^w(r,z,t)$ given by equation (\ref{eq:HOMO_radius_velocity_step1}) is

        \begin{equation}\label{eq:HOMO_radial_velocity}
        v_r^w(r,z,t) = \frac{3 r v_e^w}{4 h_0(t)h(r,t)^2} \left(2h(r,t)z - z^2\right).
        \end{equation}

        Now, we calculate the vertical component of the velocity field $v_z^w(r,z,t)$.
        From equations (\ref{eq:HOMO_continuity}) and (\ref{eq:HOMO_radial_velocity}), we have
        \begin{equation}\label{eq:HOMO_vz}
v_z^w(r,z,t) = \frac{v_e^w}{2} \left( \frac{z^3  }{h(r,t)^3} \left( 1+\frac{r^2}{R^2} \right) -\frac{3z^2}{h(r,t)^2}\right)
        \end{equation}
        where we used the boundary condition $v_z^w(r, z=0, t) = 0$.

        \subsubsection{Evaporation with flux singularity at the contact line for a dry surface}

        To compare this result with the configuration of a drop on a dry surface, we now consider the flux singularity at the drop contact line.
        From equations (\ref{eq:HOMO_temporal_derivative}) and (\ref{eq:DEE_Q}), we have

        \begin{equation}
            \frac{\partial h}{\partial t} = -  \frac{8{\cal D} (c_s - c_\infty) }{\pi \rho R} \left( 1 - \frac{r^2}{R^2}\right).
        \end{equation}
        and the mean radial velocity can be derived from (\ref{eq:bilan_local_end}),

        \begin{multline}
            \bar{v}_r^d(r,t) = \frac{2{\cal D} (c_s - c_\infty)}{\pi \rho} \frac{R  }{rh(r,t)} \\ \left( \left(1 - \frac{r^2}{R^2}\right)^{1/2} - \left(1 - \frac{r^2}{R^2}\right)^2\right)\label{eq:DEE_radial_velocity_average_simplified}
        \end{multline}
        with the boundary condition $\bar{v}_r^d(r, z=0, t) = 0$.
        Therefore, we obtain for the horizontal velocity the equation (\ref{eq:DEE_radial_velocity}) below and, from the continuity equation (\ref{eq:HOMO_continuity}), the vertical velocity given by equation (\ref{eq:DEE_vz}):

        \begin{strip}
        \begin{equation}
            v_r^d(r,z,t) = -  \frac{3{\cal D} (c_s - c_\infty)}{\pi \rho h(r,t)^3} \frac{R}{r}  \left( \left( 1 - \frac{r^2}{R^2} \right)^{1/2} - \left(1 - \frac{r^2}{R^2}\right)^2\right) (z^2 - 2h(r,t)z).\label{eq:DEE_radial_velocity}
        \end{equation}
        \end{strip}

\begin{strip}
        \begin{multline}\label{eq:DEE_vz}
            v_z^d(r,z,t) = \frac{3{\cal D} (c_s - c_\infty)}{\pi \rho R h(r,t)^3} \left[ \frac{2h_0 }{h} (z^3 - 2hz^2)
                \left( \left(1-\frac{r^2}{R^2}\right)^{1/2} - \left(1 - \frac{r^2}{R^2}\right)^2\right)\right. \\
                + \left. \left( \frac{z^3 - 3hz^2}{3} \right) \left(4 \left(1-\frac{r^2}{R^2}\right) -
            \left( \frac{1}{1-\frac{r^2}{R^2}} \right)^{1/2} \right)  \right].
        \end{multline}
\end{strip}
        \subsubsection{Typical flow fields}

        \begin{figure*}
            \centering
            \includegraphics[width=.47\linewidth]{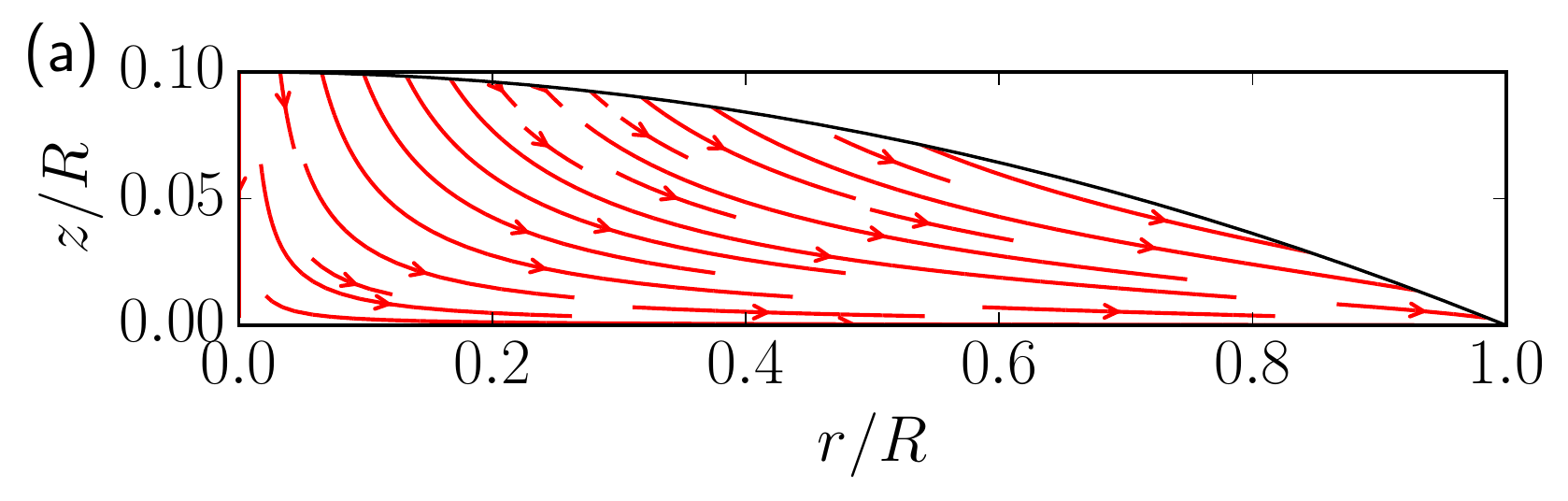}
            \includegraphics[width=.47\linewidth]{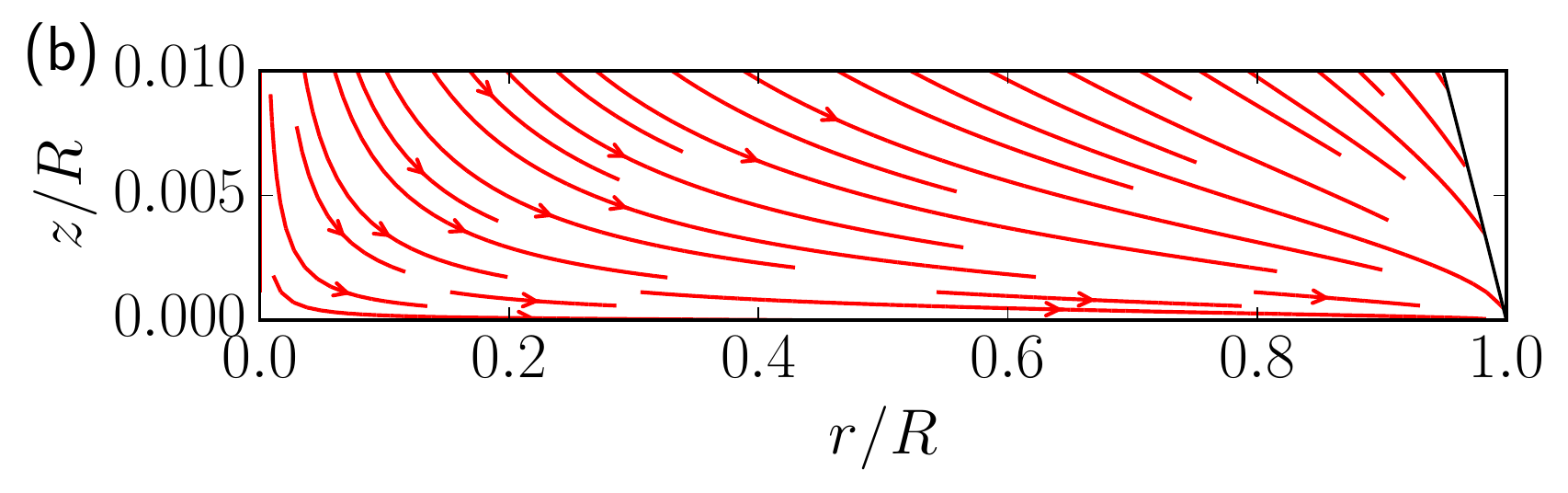}\\
            \includegraphics[width=.47\linewidth]{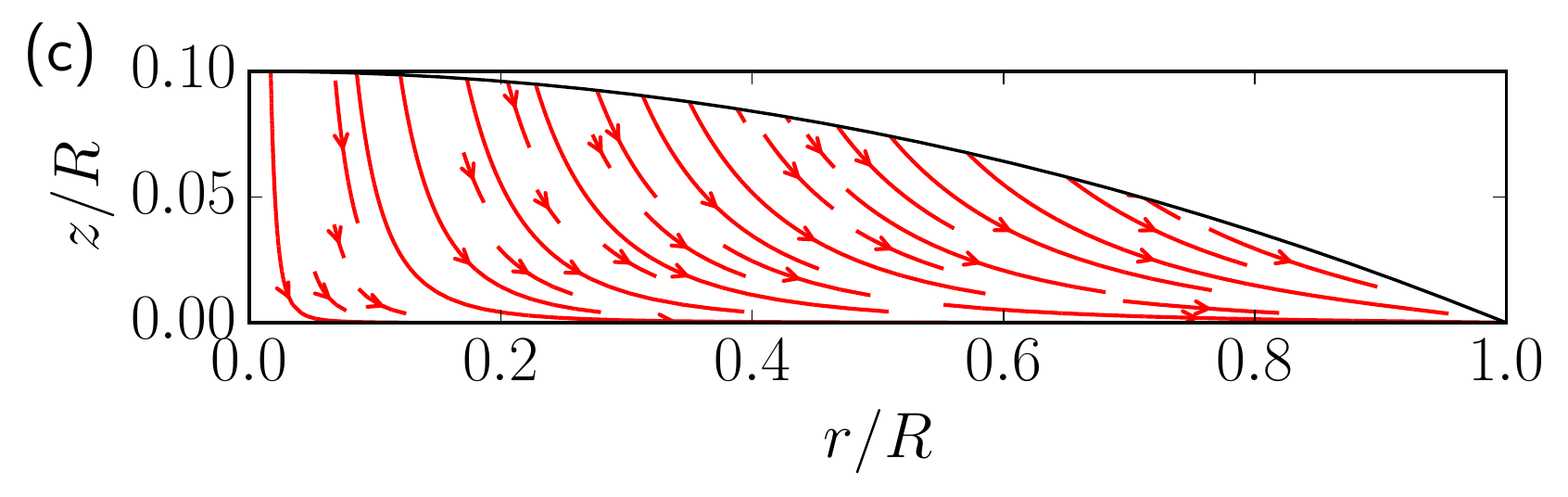}
            \includegraphics[width=.47\linewidth]{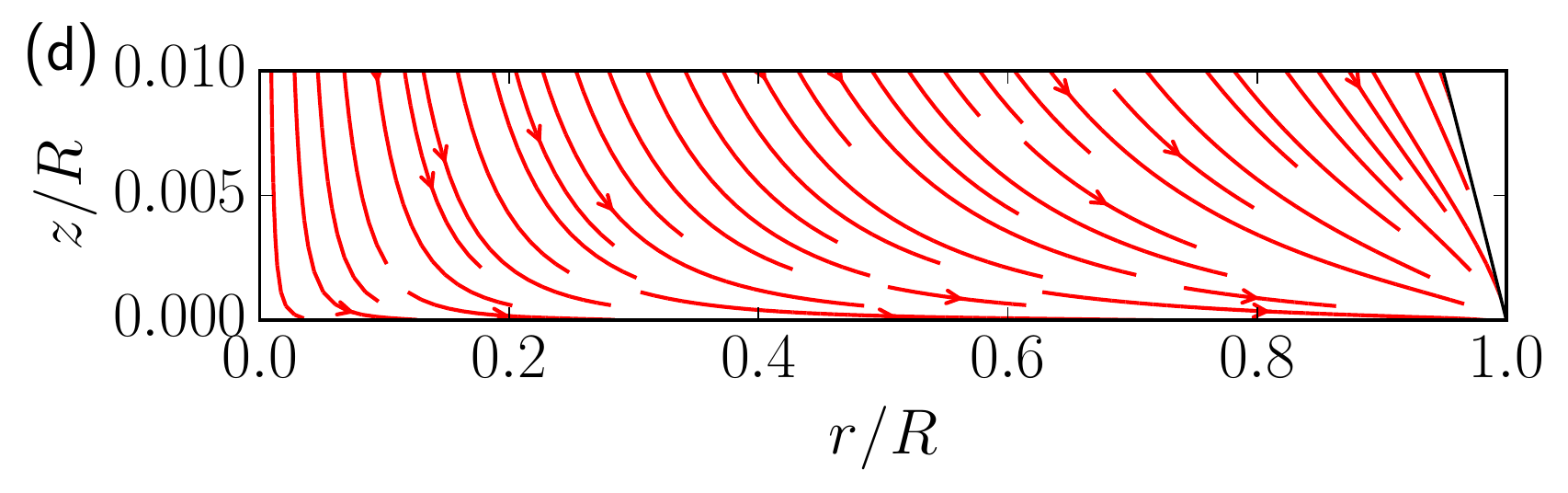}\\
            \includegraphics[width=.47\linewidth]{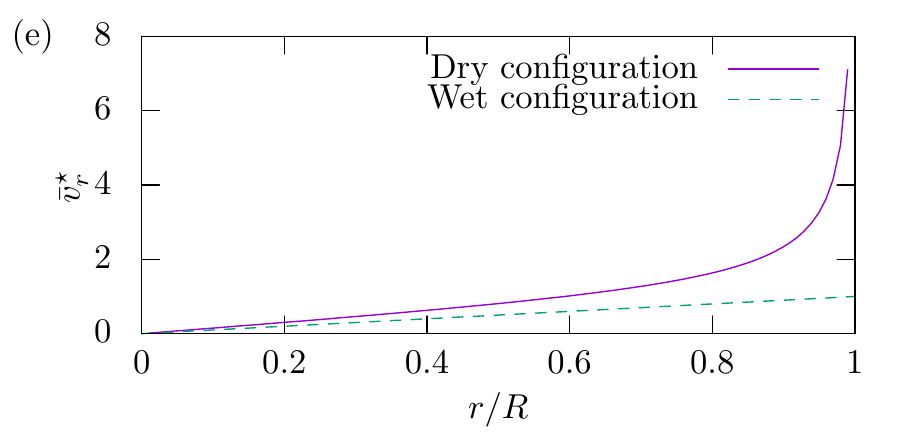}
            \includegraphics[width=.47\linewidth]{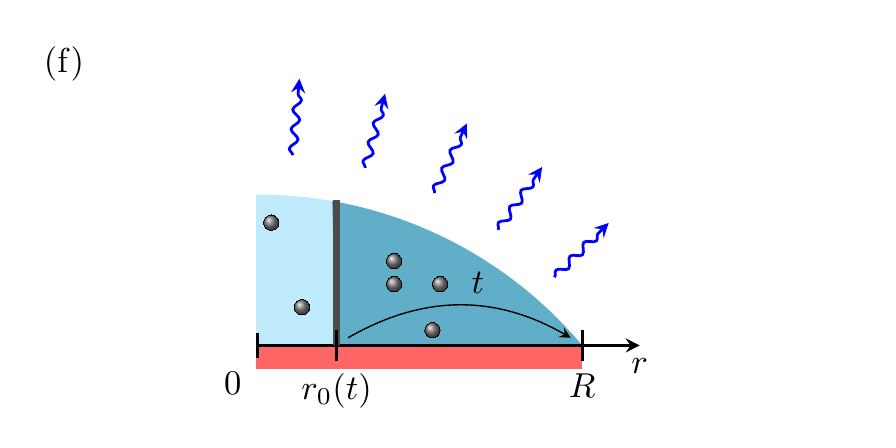}
            \caption{
                Streamlines corresponding to the velocity fields
                (a-b) for the wet configuration, given by equations (\ref{eq:HOMO_radial_velocity}) and (\ref{eq:HOMO_vz}), and
                (c-d) for the dry configuration,
                given by equations (\ref{eq:DEE_radial_velocity}) and (\ref{eq:DEE_vz}).
                The drop aspect ratio is $h_0(t)/R=0.1$.
            The black solid line represents the liquid-vapor interface.
                (e) Mean radial dimensionless velocity $\bar v_r^\star$
                defined as
                $\bar v_r^w / (v_e^w R/(2h_0(t)))$ for the wet configuration and $ \bar v_r^d / (2{\cal D} (c_s - c_\infty)/(\pi \rho h_0(t)))$ for the dry configuration with $h_0(t)/R=0.1$.
                (f) Schematic that illustrates the principle of the calculation for the number of particles accumulated at the contact line.
                The particles contained in the dark blue area at $t=0$ are transported to the contact line at time $t$.
        }\label{fig:streamlines_deegan}
        \end{figure*}

        To illustrate the results from the previous two sections, we present in Figure \ref{fig:streamlines_deegan}(a-d) the streamlines for both configurations.
        In Figure \ref{fig:streamlines_deegan}(e), we plot the average radial velocity as a function of the radial position.
        The average radial velocity diverges near the contact line in the dry configuration because of the divergence of the evaporative flux.
        In the wet configuration, this divergence does not exist and the radial velocity increases linearly along the radial position.

        %%%%%%%%%%%%%%%%%%%%%%%%%%%%%%
        % particle transport
        %%%%%%%%%%%%%%%%%%%%%%%%%%%%%%

        \subsection{Particle transport}

        In this section, we establish the time evolution of the number of particles accumulated at the contact line \cite{Deegan2000,Popov2005,Monteux2011,Berteloot2012a}.
        These particles are advected by the radial flow, which has a velocity $v_r(r,z,t)$, equations (\ref{eq:HOMO_radial_velocity} and (\ref{eq:DEE_radial_velocity}).
        In the following,  we assume that the particle concentration is independent of the $z$-coordinate and so we consider the average radial velocity $\bar{v}_r(r,t)$.
        Thus, the radial position $r(t)$ of a narrow slice of liquid is given by
        \begin{equation}\label{eq:pos_veolocity}
            \bar{v}_r(r(t),t) = \frac{ \mbox{d} r }{ \mbox{d} t}.
        \end{equation}
        We denote  $r_0(t)$ the radial position for which particles initially present at this position at $t=0$ reach the contact line at time $t$.
        This position can be calculated by integrating equation (\ref{eq:pos_veolocity}).

        The number of particles $N_{cum}(t)$ accumulating at the contact line is the sum of the particles contained in the volume between $r_0(t)$ and $R$ (Fig.~\ref{fig:streamlines_deegan}(f)).
        \begin{equation}
            N_{cum}(t) = 2\pi \int_{r_0(t)}^R c_0  h(r', t=0)\, r' \mbox{d}r',\label{eq:number_particles}
        \end{equation}
        where $c_0$ is the initial particle concentration.
        Therefore,  equation (\ref{eq:number_particles}) with the drop shape $h(r,t)$ given by equation (\ref{eq:HOMO_drop_shape}) becomes
        \begin{equation}\label{eq:number_particles_dry}
            N_{cum}(t) = 2\pi c_0 h_i R^2 \left(\frac{1}{4} - \frac{1}{2} \frac{r_0(t)^2}{R^2} + \frac{1}{4}  \frac{r_0(t)^4}{R^4}\right),
        \end{equation}
        which depends on the radial velocity through $r_0(t)$, which has to be determined, and thus on the evaporative conditions.
        The initial drop height is $h_i = h(r=0, t=0)$.

        We consider first the wet configuration.
        Using equations (\ref{eq:HOMO_vr_bar_final}) and (\ref{eq:HOMO_height_time}) in equation (\ref{eq:pos_veolocity}), we have
        \begin{equation}\label{eq:r0_dry}
            \frac{r_0^w(t)}{R} = \left(1-\frac{t}{\tau_e^w}\right)^{1/4}.
        \end{equation}
        Thus, the number of particles accumulated at the contact line can be simply calculated by using equations (\ref{eq:number_particles_dry}) and (\ref{eq:r0_dry}).

        Now, we consider the dry configuration.
        The time evolution of the drop height follows equation (\ref{eq:HOMO_height_time}) with an evaporation time $\tau_e^d = \frac{\rho h_i R \pi}{8{\cal D} (c_s - c_\infty)}$.
        With equations (\ref{eq:DEE_radial_velocity_average_simplified}) and (\ref{eq:pos_veolocity}), we have
        \begin{equation}
            \frac{X}{\sqrt{X} - X^2} \, {\rm d}X =  \frac{1}{2} \frac{1}{(1-t/\tau_e^d)} \, {\rm d} (t/\tau_e^d),
        \end{equation}
        where $X = 1 -(r_0^d/R)^2$.
        Thus, after integration, we obtain
        \begin{equation}\label{eq:r0_wet}
            \frac{r_0^d(t)}{R} = \sqrt{ 1 - \left(1-\left(1-\frac{t}{\tau_e^d}\right)^{3/4}\right)^{2/3}}.
        \end{equation}
        A more general solution taking into account the width of the ring has been found previously by Popov \cite{Popov2005}.
        We recover this solution in the limit of a thin ring observed for dilute suspensions.
        From equation (\ref{eq:number_particles_dry}) with equations (\ref{eq:r0_dry}) or (\ref{eq:r0_wet}), we have a prediction for the time evolution of the number of particles $N_{cum}$ accumulated at the contact line for both evaporating conditions.
        In the next section, we compare these predictions to experimental observations.

        %%%%%%%%%%%%%%%%%%%%%%%%%%%%%%
        %
        % PROTOCOL
        %
        %%%%%%%%%%%%%%%%%%%%%%%%%%%%%%

        \section{Experiments}\label{sec:experiments}
        \subsection{Materials and methods}

\begin{figure}
            \centering
            \includegraphics[width=\linewidth]{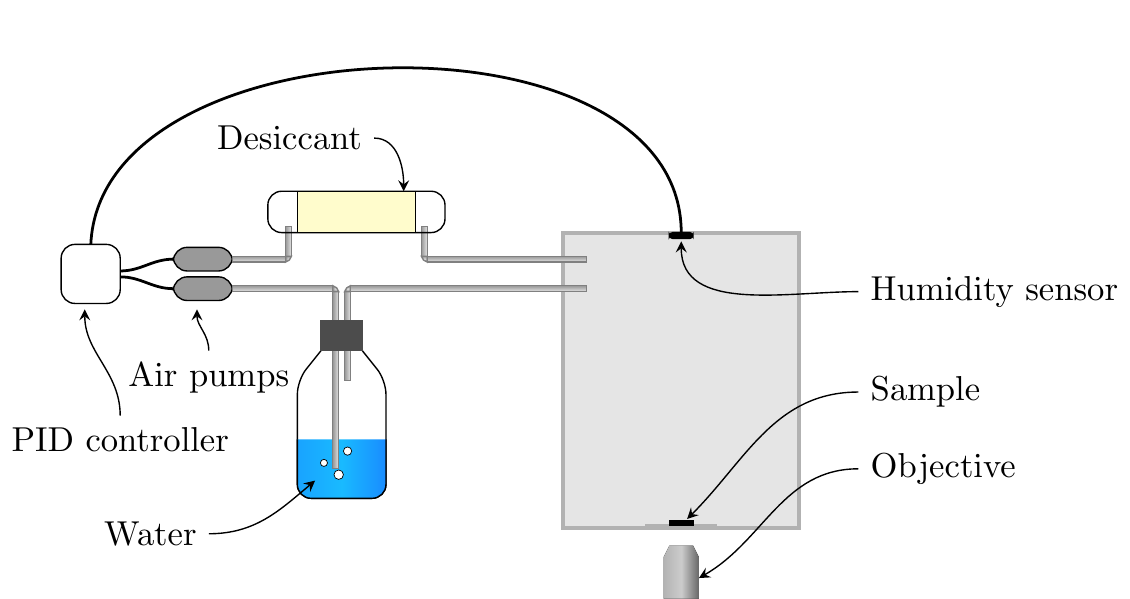}
            \caption{
                Schematic of the experimental setup.
                A PID controller injects dry or moist air into the chamber based on the relative humidity measured by the sensor.
            }\label{fig:setup}
        \end{figure}

        Our experiments of controlled drying of drops are performed in a home-made glove box ($30\times30\times30$ cm$^3$), which has humidity regulation based on an Arduino and a humidity sensor placed far from the drying drop (Figure \ref{fig:setup}).
        Dry air is produced by circulating ambiant air in a container filled with dessicant made of anhydrous calcium sulfate (Drierite) and moist air is obtained by bubbling air in water.
        The relative humidity is set to $50$\% in all of our experiments.
        The glove box is placed on the top of an inverted fluorescence microscope (Leica DMI4000 B) equipped with a 4$\times$ objective (Plan achromatic, Olympus) and a Hamamatsu camera (digital camera ORCA-Flash4.0 C11440).
        The acquisition is automated with the software Micromanager \cite{Edelstein2010}.

        Colloidal suspensions consist of fluorescent carboxylate-modified particles of $1$ $\mu$m diameter (Lifetechnologies), which are suspended in deionized water at a concentration of 2000 particles/mm$^3$, which corresponds to a volume fraction of $\phi=8\times 10^{-6}$.
        The suspension is surfactant free and no Marangoni effect is observed during the experiment.
        The surface tension measured with a pendant drop technique is $\gamma = 70\pm1$ mN/m.
        For all drop evaporation experiments, the substrate consists of a circular glass disk of diameter $2R=3$ mm and of thickness 0.15 mm (Thomas Scientific).
        Prior to each experiment, a new glass disk is placed in a plasma chamber for 5 min to enhance its wettability.
        A drop of $1.5$ $\mu\ell$ is dispensed with a micropipette (Research Plus, Eppendorf) on the glass disk.
        This volume is chosen to fully cover the glass disk with the colloidal suspension and it ensures that all the drops have the same initial size and shape.

        Two configurations are investigated, dry and wet surfaces, for the drop evaporation experiments; the configurations effectively change the local humidity of the drop's environment between the two extremes.
        For experiments with dry surfaces, to conveniently deposit a drop on the glass disk, the disk is placed on a 3 mm thick PDMS slab bonded to a glass slide, which holds the disk in place by adhesion.
        In contrast, a humid surface is made of a polydimethyl acrylamide (PDMA) hydrogel \cite{Sudre2011}.
        The composition of the gel is set by $m_{mono} / (m_{mono} + m_{w}) = 0.1$ and $[\rm{MBA}]/[\rm{mono}] = 2\times 10^{-2}$, where $m_{mono}$ and $m_{w}$ are, respectively, monomer (N,N-dimethylacrylamide) and water masses and $[\rm{MBA}]$, $[\rm{mono}]$ are, respectively, molar concentrations of the crosslinker (N,N'-methylene-bis-acrylamide) and the monomer.
        The quantity of initiator (potassium persulfate and N,N,N',N'-tetramethylethylenediamine) is set to a molar ratio of 1\% of the monomer quantity.
        All chemicals are purchased from Sigma-Aldrich, USA.
        The solution is poured in a mold made of two glass plates ($75\times50$ cm$^2$, Dow Corning) separated with a rubber spacer of $3$ mm thickness (McMaster-Carr).
        After three days, the gel is removed from the mold, swollen in deionized water overnight and its surface is gently wiped to remove any excess of water.
        A hydrogel disk of $3.5$ cm diameter is prepared and a glass disk is placed at its center.

        Fluorescence images are taken in the plane of the glass disk surface every $1.5$ s for the dry surface and every $3$ s for the wet surface.
        Particle trajectories detected on fluorescence images are calculated by using the library Trackpy  \cite{Allan2014}.
        These trajectories are calculated on sequences of $6$ and $20$ images for dry and wet surfaces, respectively.

        The data analysis is performed with the programming language Python, the scientific stack numpy-scipy \cite{Oliphant2007,Millman2011,Walt2011} and Pandas \cite{McKinney2010}.
        The detection of the position of the glass disk  on a brightfield image is performed with Scikit-image by using a Canny filter and the circular Hough transform \cite{Vanderwalt2014}.

        Experimentally, we observe that the particles have an outward radial motion.
        Thus, to evaluate the time evolution of the number of particles accumulated at the contact line, we use the particle trajectories.
        For each time sequence, we consider the trajectories that cross a circle centered on the glass disk with a radius $0.92\,R$.
        We count as positive the trajectories that have an outward direction and negative the inward trajectories.
        In both experimental conditions studied in this work, the flow is outward and inward trajectories can be observed on short timescales because of Brownian motion.

        %%%%%%%%%%%%%%%%%%%%%%%%%%%%%%
        %
        % OBSERVATIONS / DISCUSSION
        %
        %%%%%%%%%%%%%%%%%%%%%%%%%%%%%%

        \subsection{Experimental observations and discussion} \label{sec:observation}

        We designed experiments to track the motion of particles, and their corresponding accumulation at the contact line as described in the previous section.
        From the particle trajectories calculated by particle tracking, we are able to count the particles that are approaching the contact line, and to measure their velocities.

        \begin{figure*}
            \centering
            \includegraphics[width=.45\linewidth]{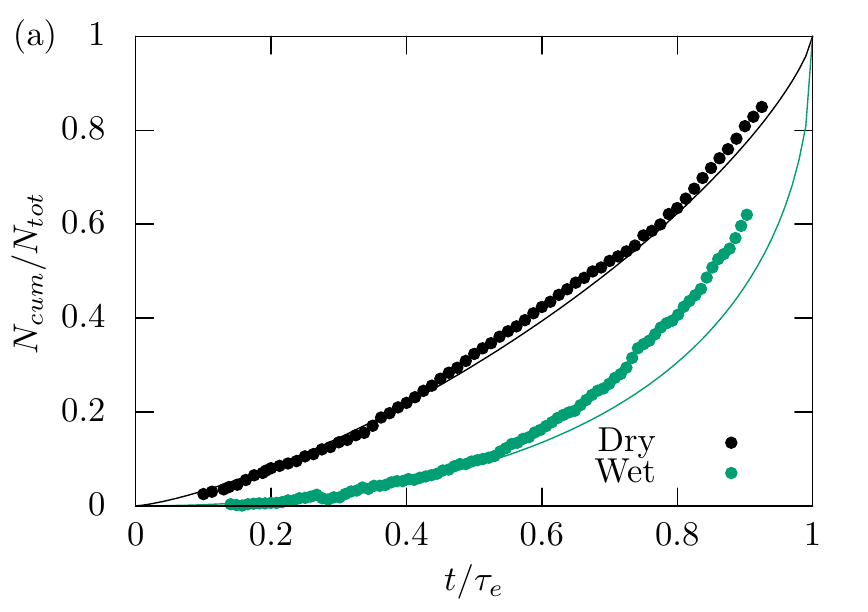}
            \includegraphics[width=.45\linewidth]{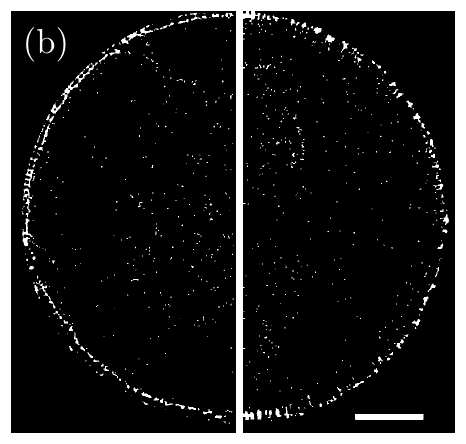}
            \caption{
                (a) Time evolution of the number of particles accumulated at the contact line for dry and wet configurations.
                The time is non-dimensionalized by the time $\tau_e$ at which the contact line recedes.
                The solid lines represent the equation (\ref{eq:number_particles_dry}) for $N_{cum}(t)$ for $r_0^d(t)$ and $r_0^w(t)$, respectively, given by equations (\ref{eq:r0_dry}) and (\ref{eq:r0_wet}).
                (b) Final deposit for dry (left) and wet (right) configurations. The scale bar represents 500 $\mu$m.
            }\label{fig:result1}
        \end{figure*}

        In Figure \ref{fig:result1}(a), we show for both dry and wet configurations, the time evolution of the  particles accumulated at the contact line.
        These values are non-dimensionalized by the total number of particles known from the particle concentration and the experimental evaporation time is normalized by the time at which the contact line starts to recede.

        In the theoretical model, all the particles are assumed to reach the contact line since there is a radial flow that replenishes the contact line during the constant radius evaporation mode (Figure \ref{fig:streamlines_deegan}).
        Indeed, the vertical velocity $v_z$ is always zero at $z=0$, which means that the particles can reach the solid surface only through Brownian motion, except in the vicinity of the contact line where the liquid height is comparable to the particle diameter.
        Experimentally, the particles can be trapped before the end of the evaporation when the liquid film has a thickness of the order of magnitude of the particle size.
        In fact, some particles do not reach the contact line and are deposited in the inner part of the wetted area as illustrated in Figure \ref{fig:result1}(b).
        Nevertheless, we observe that changing the evaporation conditions does not affect significantly the quantity of particles forming the ring.

  The percentage of particles that do not reach the drop edge is about $12$\% and $14$\% for the dry and wet configurations, respectively. These particles can originate either because they did not reach the drop edge before the contact line recedes or because of sedimentation. However, since the number of counted particles is not significantly different for the two drying times $\tau_e^d$ and $\tau_e^w$, sedimentation can be neglected.

        As displayed in Figure \ref{fig:result1}(a), the theoretical predictions based on equation (\ref{eq:number_particles_dry}) are in good agreement with our experimental observations.
        The agreement for the dry case confirms previous results obtained by Deegan \cite{Deegan2000}, Popov \cite{Popov2005} and later by Berteloot \textit{et al.} \cite{Berteloot2012a}, among others, whereas the agreement of theory and measurements for the wet configuration validates our choice of a homogeneous evaporative flux at the liquid-vapor interface.
        A more accurate prediction would probably require a more detailed description of the evaporative flux at the surface of the drop.
        We conclude that the dynamics of the particle accumulation at the contact line strongly depends on the spatial variation of the evaporative flux but does not affect significantly the final ring pattern.

\begin{figure}[h]
            \centering
            \includegraphics[height=6.5cm]{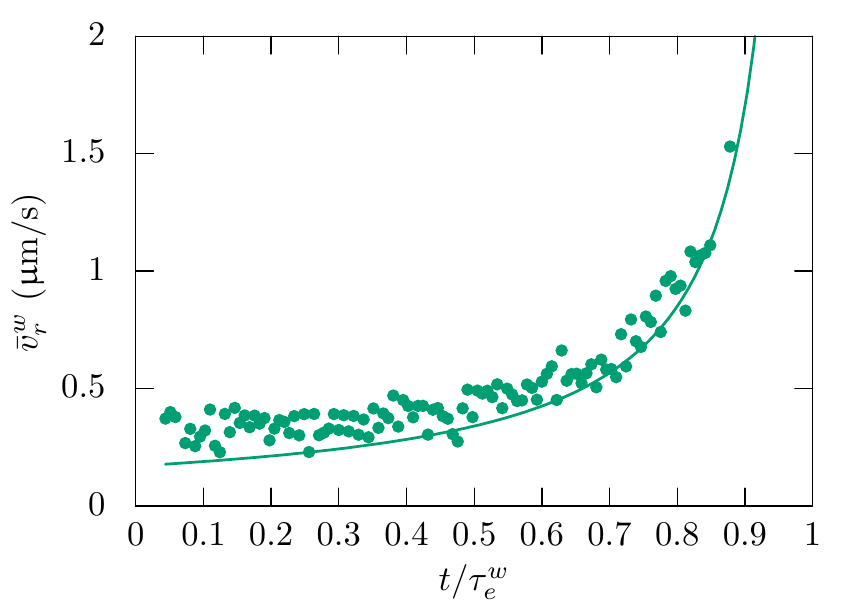}
            \caption{
  				Mean radial velocity $\bar v_r^w(r,t)$ as a function of the dimensionless time $t/\tau_e^w$.
                The radial position corresponds to $r\approx 0.9 R$.
            }\label{fig:v_r_bar_with_time}
        \end{figure}

In the dry configuration, equation (\ref{eq:DEE_radial_velocity_average_simplified})
is combined with equation (\ref{eq:DEE_height_time}) and we use the definition of the drying time $\tau_e^d$ to obtain
\begin{equation}\label{eq:DEE_v_r_bar_tau}
    \bar v_r^d (r,t) = \frac{r}{4 \tau_e^d} f\left(\frac{r^2}{R^2}\right) \left(\frac{1}{1-t/\tau_e^d}\right),
\end{equation}
where $f$ is a function of $r^2/R^2$.
This equation shows that the average radial velocity $\bar v_r^d$ diverges in time for $t=\tau_e^d$.
This feature has been indicated  by Gelderblom \textit{et al.} \cite{Gelderblom2012} for drops in a dry configuration and we show that the same mechanism is obtained in the wet configuration.
The corresponding equation to equation (\ref{eq:DEE_v_r_bar_tau}) for the wet configuration is
\begin{equation}\label{eq:HOMO_v_r_bar_tau}
    \bar v_r^w (r,t) = \frac{r}{4 \tau_e^w} \left(\frac{1}{1-r^2/R^2}\right) \left(\frac{1}{1-t/\tau_e^w}\right).
\end{equation}
In figure \ref{fig:v_r_bar_with_time}, the comparison of the mean radial velocity deduced from the particle tracking with the theoretical prediction given by equation (\ref{eq:DEE_v_r_bar_tau}) shows good agreement.

        We must also compare the predictions for the drying times with the experimental observations.
        These times are $\tau_e^d\approx 1000$ s and $\tau_e^w\approx8000$ s for the dry and wet configurations, respectively.
        In the dry configuration, the mass Grashof number (equation (\ref{eq:grashof_number})) for the drop is ${\rm Gc} = 0.7$.
        From the diffusion-limited model, the prediction $\tau_e^d = \frac{ \rho h_i R \pi}{8{\cal D} (c_s - c_\infty)} \approx 1500$ s is close to the observed value.
        The small discrepancy can be mainly explained by the assumption of a small contact angle $\theta_c$ for the evaporative flux in the model, which overestimates the evaporative flux at the beginning of the experiment.
       Stauber \textit{et al.} \cite{Stauber2014} provide an expression of the constant radius duration including the effect of the time evolution of the contact angle.
        We can numerically estimate this time to be $\approx 1150$~s.
        For the wet configuration, the Grashof number for the hydrogel disk is ${\rm Gc} = 10^3$, which highlights the importance of buoyant forces and convection in the vapor phase.
        Thus, as expected, the prediction of the drying time from diffusion-limited model $\tau_e^w = \frac{h_i }{2 v_e^w} \approx34000$ s, is much larger than the experimental value.

        Recent studies \cite{Kelly-Zion2011,Carle2013a} investigated the effect of convection on the evaporation of highly volatile droplets and they rationalized their findings with the mass Grashof number.
        For mass Grashof numbers between $5$ to $10^5$, they observed experimentally that the total evaporation velocity ${\cal V}^w$ can be written in the form
        \begin{equation}\label{eq:pheno}
            {\cal V}^w  = v_e^w (1 + v_c^\star({\rm Gc})),
        \end{equation}
        where $v_e^w$ is the diffusion-limited evaporation velocity and $v_c^\star({\rm Gc})$ is a dimensionless velocity, which is a power law of the Grashof number, $v_c^\star \propto {\rm Gc}^\beta$ with $\beta \approx 0.2$ \cite{Kelly-Zion2011,Carle2013a}.
        Therefore, from expression (\ref{eq:pheno}) and their phenomenological parameters, we can estimate ${\cal V}^w \in [$3.5, 4.5$] v_e^w$.
        Thus, the corrected theoretical prediction for the drop lifetime in wet conditions $\tau_e^w $ is between  $7500$ and $9 000$ s, which is in reasonable agreement with our observations.
        Consequently, the drying of a hydrogel disk cannot be considered to be limited by the diffusion of vapor in air.
        The convection accelerates the drying velocity of the drop.

        Finally, we would like to stress that the diffusion-limited evaporative conditions can be relevant for some applications.
        Indeed, different solvent properties or smaller systems such as a surface covered with water droplets of few tens of micrometer of diameter can evaporate with negligible convection effects.

        %%%%%%%%%%%%%%%%%%%%%%%%%%%%%%
        %
        % CONCLUSION
        %
        %%%%%%%%%%%%%%%%%%%%%%%%%%%%%%

        \section{Conclusion}

        In this paper, we have studied the drying dynamics of drops containing colloidal microspheres in a dilute regime in two drying configurations.
        The first configuration corresponds to a sessile drop dispensed on a dry flat surface and is well-studied in the literature.
        As the evaporation proceeds, particles migrate toward the contact line and form a solid ring of packed particles that pins the contact line.
        The radial flux is the consequence of the contact line pinning to compensate the loss of solvent.
        However, the details of the radial velocity depends on the evaporation profile at the drop interface.
        For the configuration of a drop on a dry surface, the evaporative flux diverges at the contact line.

        We explored the ``wet'' configuration of a drop placed at the center of a drying surface.
        Theoretically, depending on the mass Grashof number, the evaporation can be limited by the diffusion of vapor in the air or dominated by convection.
        For diffusion-limited evaporation, we established that the evaporative flux at the surface of the drop is homogeneous.
        When convection is dominant, a simple analysis of the air flow above the center of the disk indicates that the evaporative flux has a variation along the radius, which is much smaller than the divergence observed for a droplet on a dry surface.
        We concluded that, as a first approximation, a homogeneous evaporative flux can be also applied as a boundary condition for convection-dominated evaporation.
        Thus, we derived the velocity profiles in the drop for a homogeneous evaporation speed and we calculated the time dependence of the number of particles accumulated at the contact line.

        We compared experimentally the effect of the dry and wet conditions  on the dynamics of the particle accumulation under these two evaporating conditions.
        Our observations are in good agreements with the two respective models, which validates our approach regarding the homogeneous evaporative flux for the wet configuration.
        Also, as predicted by our model, we did not observe a significant variation of the final particle number density at the ring depending on the evaporating conditions.
        Finally, the evaporation timescales confirm that a millimeter size sessile droplet evaporates according to a diffusion-limited model.
        However, for a droplet in the center of a centimeter size evaporating gel, convection above the disk dominates as suggested by the large mass Grashof number.
        This significantly reduces the drop lifetime compared to the prediction of pure-diffusive evaporation.
        As we indicated that convection can be important for the drop evaporation surrounded by an evaporative surface, future works should treat more precisely the effect of convection to describe more accurately the drying conditions.

        \section*{Acknowledgments}
        F.B. acknowledges that the research leading to these results received funding from the People Programme (Marie Curie Actions) of the European Union's Seventh Framework Programme (FP7/2007-2013) under REA grant agreement 623541.
        We thank J. Dervaux and L. Limat for helpful discussions.

        %%%%%%%%%%%%%%%%%%%%%%%%%%%%%%
        %
        % Appendix
        %
        %%%%%%%%%%%%%%%%%%%%%%%%%%%%%%

        \section{Appendix}\label{sec:Appendix}

\begin{figure}
            \centering
            \includegraphics[height=6.5cm]{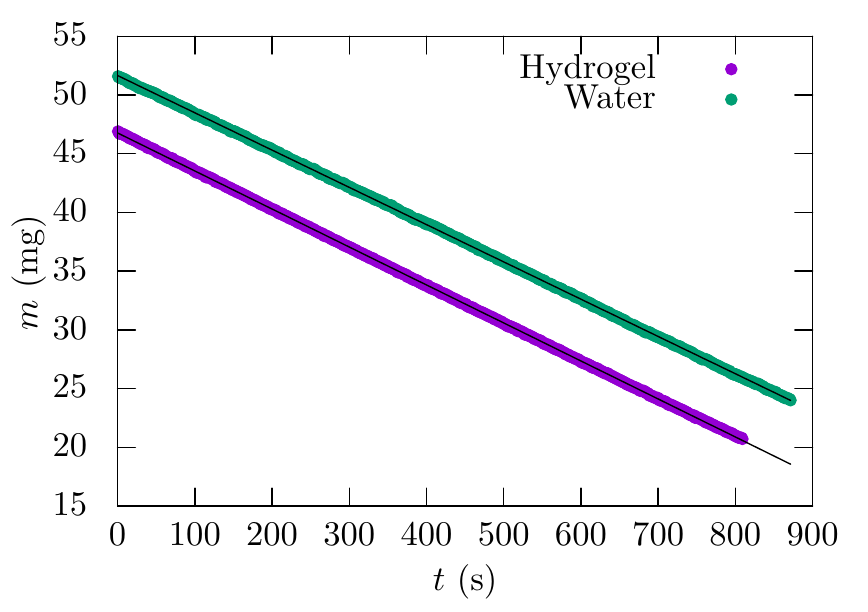}
            \caption{
                Time evolution of the weight of hydrogel disk of radius $R_g=1.75$ cm and water in a short cylindrical container of the same radius.
                The initial weight is arbitrarily shifted to distinguish the data points.
                Black solid lines are linear fits with a slope $-0.032\pm 0.001$ mg/s.
                Measurements are done in a controlled atmosphere at a relative humidity $R_H=50$ \%.
            }\label{fig:appendix1}
        \end{figure}

        In the main text, we assumed that a hydrogel dries at the same speed as pure water.
        In order to validate this assumption, we measured the drying velocity of water and gel in the same environmental conditions.
        The drying kinetics of a water reservoir and a gel slab of circular shape (radius $R_g=1.75$ cm) are recorded by placing them on a scale in a glove box at $50$\% relative humidity.
        The evolution of the mass versus time is plotted in Figure \ref{fig:appendix1}.
        The weights decrease linearly with time at the same rate for the water and the gel.
        These measurements validate the assumption that the drop deposited in the center of the piece of gel is drying with the velocity given by equation (\ref{eq:HOMO_avvelocity}).

        The total evaporative flux of water from the gel $\mathcal{Q}_e$ is given by the integration of equation (\ref{eq:HOMO_current}) over the total surface of the gel,
        \begin{equation}
            \mathcal{Q}_e = \int_0^{R_g} v_e^w \pi r {\rm d}r =  \frac{4 {\cal D} (c_s - c_\infty) R_g}{\rho}.
            \label{eq:HOMO_flux}
        \end{equation}
        From the linear fit of data presented in Figure \ref{fig:appendix1}, we can deduce the total evaporative flux
        \begin{equation}
            {\cal Q}_e = \frac{1}{\rho}\frac{\mbox{d}m}{\mbox{d}t} = 3.2\times 10^{-11}~\textrm{m}^3/\textrm{s}.
        \end{equation}
        The theoretical value defined by equation (\ref{eq:HOMO_flux}) gives ${\cal Q}_e = 1.4 \times 10^{-11}~\textrm{m}^3/\textrm{s}$.
        Thus, the experimental average flux is about a factor of two larger than the theoretical value, which supports a convection enhanced evaporation.

        \bibliography{biblio}
        \bibliographystyle{unsrt}

        \end{document}